\documentclass{article}

\begin{document}

\title{P, C and T for Truly  Neutral Particles}
%\thanks{Presented at the QTRF-5, V\"axj\"o. Sweden, June 14-18, 2009  and at the ICSSUR09, %Olomouc, Czech Republic, June 22-26, 2009.}}

\author{Valeriy V. Dvoeglazov\\
Universidad de Zacatecas\\Ap. Postal 636, Suc. 3 Cruces, C. P. 98064\\Zacatecas, Zac., M\'exico}

\date{\empty}

\maketitle

\begin{abstract}
We  present a realization of a quantum field theory, envisaged many years ago by Gelfand, Tsetlin, Sokolik and  
Bilenky. Considering the special case of the $(1/2,0)\oplus (0,1/2)$ field and developing the Majorana construct for neutrino we show that 
a fermion and its antifermion can have the same properties with respect to the intrinsic parity ($P$) operation. The transformation laws for $C$
and $T$ operations have also been given. The construct can be applied to explanation of the present situation in neutrino physics.
The case of the $(1,0)\oplus (0,1)$ field is also considered.
\end{abstract}

%\maketitle

%\newpage

During the 20th century various authors introduced {\it self/anti-self} charge-conjugate 4-spinors
(including in the momentum representation), see~\cite{Majorana,Bilenky,Ziino,Ahluwalia}. Later, Lounesto, 
Dvoeglazov, Kirchbach {\it etc} studied these spinors, they found dynamical equations, gauge transformations 
and other specific features of them. Recently, in~\cite{Kirchbach} it was claimed that ``for imaginary $C$ parities, the neutrino mass can drop out from the single $\beta$ decay trace and reappear in $0\nu\beta\beta$,...  in principle experimentally testable signature for a non-trivial impact of Majorana framework in experiments with polarized  sources" (see also Summary of the cited paper). Thus, phase factors can have physical significance in quantum mechanics. So, the aim of my talk is to remind what several researchers presented in the 90s concerning with the neutrino description.

We  define the {\it self/anti-self} charge-conjugate 4-spinors 
in the momentum space\footnote{In~\cite{Kirchbach}  a bit different notation was used referring to~\cite{Bilenky}.}
\begin{equation}
C\lambda^{S,A} (p^\mu) = \pm \lambda^{S,A} (p^\mu)\,,
C\rho^{S,A} (p^\mu) = \pm \rho^{S,A} (p^\mu)\,,
\end{equation}
where
\begin{equation}
\lambda^{S,A} (p^\mu)=\pmatrix{\pm i\Theta \phi^\ast_L (p^\mu)\cr
\phi_L (p^\mu)}\,,
\rho^{S,A} (p^\mu)=\pmatrix{\phi_R (p^\mu)\cr \mp i\Theta \phi^\ast_R (p^\mu)}\,.
\end{equation}
The Wigner matrix is
$\Theta_{[1/2]}=-i\sigma_2=\pmatrix{0&-1\cr
1&0}$\,,
and $\phi_L$, $\phi_R$ are the Ryder (Weyl) left- and right-handed 2-spinors.
The 4-spinors  $\lambda$ and $\rho$ are NOT the eigenspinors of helicity. Moreover, 
$\lambda$ and $\rho$ are NOT the eigenspinors of the parity $P=\pmatrix{0&1\cr 1&0}R$, as opposed to the Dirac case.
Such definitions of 4-spinors differ, of course, from the original Majorana definition in x-representation.
They are eigenstates of the chiral helicity 
quantum number introduced 
in the 60s, $\eta=-\gamma^5 h$.
While 
\begin{equation}
Pu_\sigma ({\bf p}) = + u_\sigma ({\bf p})\,,
Pv_\sigma ({\bf p}) = - v_\sigma ({\bf p})\,,
\end{equation}
we have
\begin{equation}
P\lambda^{S,A} ({\bf p}) = \rho^{A,S} ({\bf p})\,,
P \rho^{S,A} ({\bf p}) = \lambda^{A,S} ({\bf p})\,,
\end{equation}
for the Majorana-like momentum-space 4-spinors
on the first quantization level.

One can use the generalized form of the Ryder
relation for zero-momentum 
spinors:
\begin{equation}
\label{rbug12} \left [\phi_{_L}^h
({\bf 0})\right ]^* = (-1)^{1/2-h}\, e^{-i(\vartheta_1^L
+\vartheta_2^L)} \,\Theta_{[1/2]} \,\phi_{_L}^{-h} ({\bf 0})\,,
\end{equation}
in order to derive the dynamical equations~\cite{Dvoeglazov1}:
\begin{eqnarray}
i \gamma^\mu \partial_\mu \lambda^S (x) - m \rho^A (x) &=& 0 \,,
\label{11}\\
i \gamma^\mu \partial_\mu \rho^A (x) - m \lambda^S (x) &=& 0 \,,
\label{12}\\
i \gamma^\mu \partial_\mu \lambda^A (x) + m \rho^S (x) &=& 0\,,
\label{13}\\
i \gamma^\mu \partial_\mu \rho^S (x) + m \lambda^A (x) &=& 0\,.
\label{14}
\end{eqnarray}
These are NOT the Dirac equations (cf.~\cite{Markov}).
Similar formulation has been presented by
A. Barut and G. Ziino~\cite{Ziino}. The group-theoretical basis for such doubling has 
been first given
in the papers by Gelfand, Tsetlin and Sokolik~\cite{Gelfand} and other authors.

Hence, the Lagrangian is
\begin{eqnarray}
{\cal L}= \frac{i}{2} \left[\bar \lambda^S \gamma^\mu \partial_\mu \lambda^S - (\partial_\mu \bar \lambda^S ) \gamma^\mu \lambda^S +
\bar \rho^A \gamma^\mu \partial_\mu \rho^A - (\partial_\mu \bar \rho^A ) \gamma^\mu \rho^A +\right.\nonumber\\
\left.\bar \lambda^A \gamma^\mu \partial_\mu \lambda^A - (\partial_\mu \bar \lambda^A ) \gamma^\mu \lambda^A +
\bar \rho^S
\gamma^\mu \partial_\mu \rho^S - (\partial_\mu \bar \rho^S ) \gamma^\mu \rho^S -\right.\nonumber\\
\left. - m (\bar\lambda^S \rho^A +\bar \rho^A \lambda^S -\bar\lambda^A \rho^S -\bar\rho^S \lambda^A )
\right ]\,.
\end{eqnarray}

The connection with the Dirac spinors has been found. 
For instance~\cite{Ahluwalia,Dvoeglazov1},
\begin{eqnarray}
\pmatrix{\lambda^S_\uparrow (p^\mu) \cr \lambda^S_\downarrow (p^\mu) \cr
\lambda^A_\uparrow (p^\mu) \cr \lambda^A_\downarrow (p^\mu)\cr} = {1\over
2} \pmatrix{1 & i & -1 & i\cr -i & 1 & -i & -1\cr 1 & -i & -1 & -i\cr i&
1& i& -1\cr} \pmatrix{u_{+1/2} (p^\mu) \cr u_{-1/2} (p^\mu) \cr
v_{+1/2} (p^\mu) \cr v_{-1/2} (p^\mu)\cr}\,.\label{connect}
\end{eqnarray}
See also ref.~\cite{Gelfand,Ziino}.

It was shown~\cite{Dvoeglazov1} that the covariant derivative (and, hence, the
 interaction) can be introduced in this construct in the following way:
\begin{equation}
\partial_\mu \rightarrow \nabla_\mu = \partial_\mu - ig \L^5 B_\mu\,,
\end{equation}
where $\L^5 = \mbox{diag} (\gamma^5 \quad -\gamma^5)$, the $8\times 8$
matrix. With respect to the chiral phase transformations
the spinors retain their properties to be self/anti-self charge conjugate
spinors and the proposed Lagrangian~\cite[p.1472]{Dvoeglazov1} remains to be invariant.
This tells us that while self/anti-self charge conjugate states has
zero eigenvalues of the ordinary (scalar) charge operator but they can
possess the axial charge (cf.  with the discussion of~\cite{Ziino} and
the old idea of R. E. Marshak and others).\footnote{In fact, from this consideration 
one can recover the Feynman-Gell-Mann
equation (and its charge-conjugate equation).}
Next, because the transformations
\begin{eqnarray}
\lambda_S^\prime (p^\mu) &=& \pmatrix{\Xi &0\cr 0&\Xi} \lambda_S (p^\mu)
\equiv \lambda_A^\ast (p^\mu)\quad,\quad\\
\lambda_S^{\prime\prime} (p^\mu) &=& \pmatrix{i\Xi &0\cr 0&-i\Xi} \lambda_S
(p^\mu) \equiv -i\lambda_S^\ast (p^\mu)\quad,\quad\\
\lambda_S^{\prime\prime\prime} (p^\mu) &=& \pmatrix{0& i\Xi\cr
i\Xi &0\cr} \lambda_S (p^\mu) \equiv i\gamma^0 \lambda_A^\ast
(p^\mu)\quad,\quad\\
\lambda_S^{IV} (p^\mu) &=& \pmatrix{0& \Xi\cr
-\Xi&0\cr} \lambda_S (p^\mu) \equiv \gamma^0\lambda_S^\ast
(p^\mu)\quad
\end{eqnarray}
with the $2\times 2$ matrix $\Xi$ defined as ($\phi$ is the azimuthal
angle  related to ${\bf p} \rightarrow {\bf 0}$)
\begin{equation}
\Xi = \pmatrix{e^{i\phi} & 0\cr 0 &
e^{-i\phi}\cr}\quad,\quad \Xi \Lambda_{R,L} (0 \leftarrow
p^\mu) \Xi^{-1} = \Lambda_{R,L}^\ast (0 \leftarrow
 p^\mu)\,\,\, ,
\end{equation}
and corresponding transformations for
$\lambda^A$ do {\it not} change the properties of bispinors to be in the
self/anti-self charge conjugate spaces, the Majorana-like field operator
($b^\dagger \equiv a^\dagger$) admits additional phase (and, in general,
normalization) $SU(2)$ transformations:
\begin{equation} \nu^{ML\,\,\prime}
(x^\mu) = \left [ c_0 + i({\bf \tau}\cdot  {\bf c}) \right
]\nu^{ML\,\,\dagger} (x^\mu) \quad, \end{equation} where $c_\alpha$ are
arbitrary parameters. The conclusion
is:  a non-Abelian construct is permitted, which is based on
the spinors of the Lorentz group only (cf. with the old ideas of T. W.
Kibble and R. Utiyama) .  This is not surprising because both $SU(2)$
group and $U(1)$ group are  the sub-groups of the extended Poincar\'e group
(cf.~\cite{Ryder}).

The Dirac-like and Majorana-like field operators can
be built from both $\lambda^{S,A} (p^\mu)$ and $\rho^{S,A} (p^\mu)$,
or their combinations. 
The anticommutation relations are the following ones (due to the {\it bi-orthonormality}):
\begin{eqnarray}
[a_{\eta{\prime}} (p^{\prime^\mu}), a_\eta^\dagger (p^\mu) ]_\pm = (2\pi)^3 2E_p \delta ({\bf p} -{\bf p}^\prime) \delta_{\eta,-\eta^\prime}
\end{eqnarray}
and 
\begin{eqnarray}
[b_{\eta{\prime}} (p^{\prime^\mu}), b_\eta^\dagger (p^\mu) ]_\pm = (2\pi)^3 2E_p \delta ({\bf p} -{\bf p}^\prime) \delta_{\eta,-\eta^\prime}
\end{eqnarray}
Other (anti)commutators are equal to zero: ($[ a_{\eta^\prime} (p^{\prime^\mu}), 
b_\eta^\dagger (p^\mu) ]=0$).

In the Fock space the operations of the charge conjugation and space
inversions can be defined through unitary operators.
The time reversal operation should be defined through {\it an antiunitary}
operator. We  further assume the vacuum state to be assigned the
even $P$- and $C$-eigenvalue and, then, proceed as in ref.~\cite{Itzykson}.
As a result we have  very different properties
with respect to the space inversion operation, comparing with
the Dirac states (the case was also regarded in~\cite{Ziino}):
\begin{eqnarray}
U^s_{[1/2]} \vert {\bf p},\uparrow >^+ = + i \vert -{\bf p},
\downarrow >^+,
U^s_{[1/2]} \vert {\bf p},\uparrow >^- = + i
\vert -{\bf p}, \downarrow >^-\\
U^s_{[1/2]} \vert {\bf p},\downarrow >^+ = - i \vert -{\bf p},
\uparrow >^+,
U^s_{[1/2]} \vert {\bf p},\downarrow >^- =  - i
\vert -{\bf p}, \uparrow >^-
\end{eqnarray}
For the charge conjugation operation in the Fock space we have
two physically different possibilities. The first one,
in fact, has some similarities with the Dirac construct.
The action of this operator on the physical states are
\begin{eqnarray}
U^c_{[1/2]} \vert {\bf p}, \, \uparrow >^+ &=& + \,\vert {\bf p},\,
\uparrow >^- \,,\,
U^c_{[1/2]} \vert {\bf p}, \, \downarrow >^+ = + \, \vert {\bf p},\,
\downarrow >^- \,,\\
U^c_{[1/2]} \vert {\bf p}, \, \uparrow >^-
&=&  - \, \vert {\bf p},\, \uparrow >^+ \,,\,
U^c_{[1/2]} \vert
{\bf p}, \, \downarrow >^- = - \, \vert {\bf p},\, \downarrow >^+ \,.
\end{eqnarray}
But, one can also construct the charge conjugation operator in the
Fock space which acts, {\it e.g.}, in the following manner:
\begin{eqnarray}
\widetilde U^c_{[1/2]} \vert {\bf p}, \, \uparrow >^+ &=& - \,\vert {\bf
p},\, \downarrow >^- \,,\,
\widetilde U^c_{[1/2]} \vert {\bf p}, \, \downarrow
>^+ = - \, \vert {\bf p},\, \uparrow >^- \,,\\
\widetilde U^c_{[1/2]} \vert
{\bf p}, \, \uparrow >^- &=& + \, \vert {\bf p},\, \downarrow >^+
\,,\,
\widetilde U^c_{[1/2]} \vert {\bf p}, \, \downarrow >^- = + \, \vert {\bf
p},\, \uparrow >^+ \,.
\end{eqnarray}
This is due to corresponding algebraic structures of self/anti-self charge-conjugate spinors.
Finally, the time reversal {\it anti-unitary} operator in 
the Fock space should be defined in such a way that the formalism to be
 compatible with the $CPT$ theorem. We obtain for the $\Psi
(x^\mu)$:
\begin{eqnarray}
V^{^T} a^\dagger_\uparrow ({\bf p}) (V^{^T})^{-1} &=& a^\dagger_\downarrow
(-{\bf p})\,,\,
V^{^T} a^\dagger_\downarrow ({\bf p}) (V^{^T})^{-1} = -
a^\dagger_\uparrow (-{\bf p})\,, \\
V^{^T} b_\uparrow ({\bf p}) (V^{^T})^{-1} &=& b_\downarrow
(-{\bf p})\,,\,
V^{^T} b_\downarrow ({\bf p}) (V^{^T})^{-1} = -
b_\uparrow (-{\bf p})\,.
\end{eqnarray}

In the $(1,0)\oplus (0,1)$ representation space 
one can define the $\Gamma^5 C$ self/anti-self charge conjugate 6-component objects.
\begin{eqnarray}
\Gamma^5 C_{[1]}  \lambda ( p^\mu ) &=&\pm \lambda ( p^\mu )\,,\\
\Gamma^5 C_{[1]}  \rho ( p^\mu ) &=&\pm \rho ( p^\mu )\,.
\end{eqnarray}
The $C_{[1]}$ matrix is constructed from dynamical equations for charged spin-1 particles.
No self/anti-self charge-conjugate states are possible.
They are also NOT the eigenstates of the parity operator (except for $\lambda_{\rightarrow}$):
\begin{eqnarray}
P\lambda^S_\uparrow =+\lambda^S_\downarrow\,, P\lambda^S_\rightarrow =-\lambda^S_\rightarrow\,, P\lambda^S_\downarrow =+\lambda^S_\uparrow\,, \\
P\lambda^A_\uparrow =-\lambda^A_\downarrow\,, P\lambda^A_\rightarrow =+\lambda^A_\rightarrow\,, P\lambda^A_\downarrow =+\lambda^A_\uparrow\,.
\end{eqnarray}
The dynamical equations are
\begin{eqnarray}
\gamma_{\mu\nu} p^\mu p^\nu  \lambda^S_{\uparrow\downarrow} - m^2 \lambda^S_{\downarrow\uparrow}&=&0\,,\\
\gamma_{\mu\nu} p^\mu p^\nu  \lambda^A_{\uparrow\downarrow} + m^2 \lambda^A_{\downarrow\uparrow}&=&0\,,\\\
\gamma_{\mu\nu} p^\mu p^\nu  \lambda^S_{\rightarrow} + m^2 \lambda^S_{\rightarrow}&=&0\,,\\
\gamma_{\mu\nu} p^\mu p^\nu  \lambda^A_{\rightarrow} -m^2 \lambda^A_{\rightarrow}&=&0\,.
\end{eqnarray}
On the secondary quantization level we obtained similar results as in the spin-1/2 case.

%\bigskip

The conclusions are:
The momentum-space Majorana -like spinors are considered in the $(j,0)\oplus (0,j)$ representation space.
They have different properties from the Dirac spinors even on the classical level.
It is convenient to work in the 8-dimensional space. Then, we can impose the Gelfand-Tsetlin-Sokolik 
(Bargmann-Wight\-man-Wigner) prescription of 2-dimensional representation of the inversion group.
Gauge transformations are different. The axial charge is possible.
Experimental differencies have been recently discussed  (the possibility of observation 
of the phase factor/eigenvalue of the $C$-parity), see~\cite{Kirchbach}.
(Anti)commutation relations are assumed to be different from the Dirac case (and the $2(2j+1)$ case) due to the bi-orthonormality of the states (the spinors are self-orthogonal).
The $(1,0)\oplus (0,1)$ case has also been considered. The $\Gamma^5 C$-self/anti-self conjugate objects have been introduced. The results  are similar to 
the $(1/2,0)\oplus (0,1/2)$ representation. The 12-dimensional formalism was introduced.
The field operator can describe both charged and neutral states.


\begin{thebibliography}{99}

\footnotesize{

\bibitem[Majorana]{Majorana} E. Majorana, Nuovo  Cimento {\bf 14} (1937) 171.

\bibitem[Bilenky]{Bilenky} S. M. Bilenky and B. M. Pontekorvo, Phys. Repts {\bf 42} (1978) 224.

\bibitem[Ziino]{Ziino} A. Barut and G. Ziino, Mod. Phys. Lett. A{\bf 8} (1993) 1099; G. Ziino, Int. J. Mod. Phys. 
A{\bf 11} (1996) 2081.

\bibitem[Ahluwalia]{Ahluwalia} D. V. Ahluwalia, Int. J. Mod. Phys. A{\bf 11} (1996) 1855.

\bibitem[Lounesto]{Lounesto} P. Lounesto, {\it Clifford Algebras and Spinors.} (Cambridge University Press, 2002), Ch. 11 and 12;
R. da Roacha and W. Rodrigues, Jr., {\it Where are Elko Spinor Fields in Lounesto 
Spinor Field Classification?} Preprint math-ph/0506075.

\bibitem[Dvoeglazov1]{Dvoeglazov1} V. V. Dvoeglazov, Int. J. Theor. Phys. {\bf 34} (1995) 2467; Nuovo Cim. {\bf 108}A (1995) 1467; 
Hadronic J. {\bf 20} (1997) 435; Acta Phys. Polon. B{\bf 29} (1998) 619.

\bibitem[Dvoeglazov2]{Dvoeglazov2} V. V. Dvoeglazov, Mod. Phys. Lett. A{\bf 12} (1997) 2741.

\bibitem[Kirchbach]{Kirchbach} M. Kirchbach, C. Compean and L. Noriega, {\it Beta Decay 
with Momentum-Space Majorana Spinors.} Eur. Phys. J. A{\bf 22} (2004) 149.

\bibitem[Markov]{Markov} M. Markov, ZhETF {\bf 7} (1937) 579, 603; Nucl. Phys. {\bf 55} (1964) 130.

\bibitem[Gelfand]{Gelfand} I. M. Gelfand and M. L. Tsetlin, ZhETF {\bf 31} (1956) 1107; G. A. Sokolik, ZhETF {\bf 33} (1957) 1515.

\bibitem[Foldy]{Foldy} B. Nigam and L. L. Foldy, Phys. Rev. {\bf 102} (1956) 1410.

\bibitem[Ryder]{Ryder} L. H. Ryder, {\it Quantum Field Theory.} (Cambridge University Press, Cambridge, 1985).

\bibitem[Itsykson]{Itzykson} C. Itzykson and J.-B. Zuber, {\it Quantum Field Theory.} (McGraw-Hill Book Co., 1980), p. 156.

}

\end{thebibliography}
\end{document}